\documentclass{iopart}
\usepackage{euscript,iopams,amssymb,amsfonts,graphicx,bm}
\usepackage{pgfplots,setstack}
 \usepackage[colorlinks=true]{hyperref}
 \hypersetup{urlcolor=blue, citecolor=red}

\usepackage{float}
\usepackage{epsfig}
\usepackage{esint}
 \usepackage{tikz-cd}

\usepackage{bm,braket}

\eqnobysec

\newcommand{\calU}{{\mathcal U}}

\newcommand{\calM}{{\mathcal M}}
\newcommand{\calC}{{\mathcal C}}
\newcommand{\R}{{\mathbb R}}

\renewcommand{\P}{\mathbb{P}}
\newcommand{\PP}{\widetilde{P}}
\newcommand{\Q}{\widetilde{Q}}

\renewcommand{\e}{{\mathrm e}}
\newcommand{\E}{{\mathbb E}}

\renewcommand{\P}{\mathbb P}

\begin{document}
\title{Morphogen gradient formation in partially absorbing media}

\author{Paul C. Bressloff}
\address{Department of Mathematics, University of Utah 155 South 1400 East, Salt Lake City, UT 84112}

\begin{abstract} 
Morphogen gradients play an essential role in the spatial regulation of cell patterning during early development. The classical mechanism of morphogen gradient formation involves the diffusion of morphogens away from a localized source combined with some form of bulk absorption. Morphogen gradient formation plays a crucial role during early development, whereby a spatially varying concentration of  morphogen protein drives a corresponding spatial variation in gene expression during embryogenesis. In most models, the absorption rate is taken to be a constant multiple of the local concentration. In this paper, we explore a more general class of diffusion-based model in which absorption is formulated probabilistically in terms of a stopping time condition. Absorption of each particle occurs when its time spent within the bulk domain (occupation time) exceeds a randomly distributed threshold $a$; the classical model with a constant rate of absorption is recovered by taking the threshold distribution $\Psi(a)=\e^{-\kappa_0 a}$. We explore how the choice of $\Psi(a)$ affects the steady-state concentration gradient, and the relaxation to steady-state as determined by the accumulation time. In particular, we show that the more general model can generate similar concentration profiles to the classical case, while significantly reducing the accumulation time.

 \end{abstract}
%%%%%%%%%%%%%%%%%%%%%%%%%%%%%%%%%%%%%%%%%%%%%%%%%%%

\maketitle

\section{Introduction}

It is now well established that morphogen gradient formation plays a key role in the spatial regulation of cell differentiation during development, consistent with the French flag paradigm originally proposed by Wolpert \cite{Wolpert69,Wolpert06}. According to the French flag model, a spatially varying concentration of a morphogen protein drives a corresponding spatial variation in gene expression through some form of concentration thresholding mechanism. For example, in regions where the morphogen concentration exceeds a particular threshold, a specific gene is activated (see Fig. 1a). Hence, a continuously varying morphogen concentration can be converted into a discrete spatial pattern of differentiated gene expression across a cell population. The most common mechanism for morphogen gradient formation is thought to involve a localized source of protein production within the embryo, combined with diffusion away from the source and subsequent absorption \cite{Lander02,Ashe06,Wartlick09,Lander11,Shvartsman12,Teimouri16}. The latter is due to the binding of morphogen to membrane bound receptors and subsequent removal from the diffusing pool by endocytosis (see Fig. 1b). It follows that the the effective absorption rate depends on the rates of binding and internalization. In certain cases, the bound receptors may activate the gene expression of its cognate receptor, thus increasing the morphogen absorption rate. This results in faster absorption in regions of higher morphogen concentration \cite{Eldar03}.

\begin{figure}[b!]
\begin{center}
\includegraphics[width=12cm]{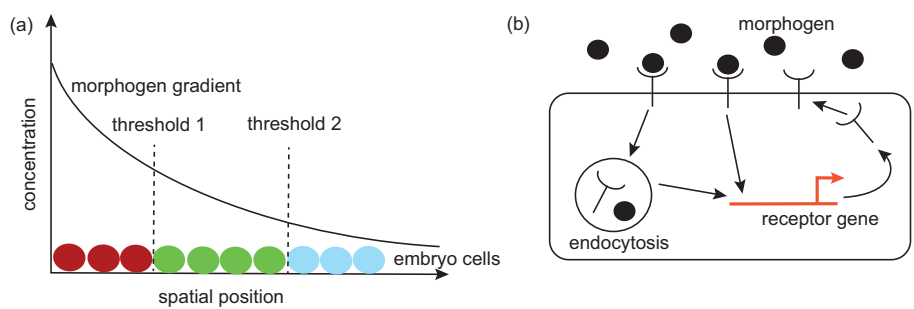}
\end{center}
\caption{A schematic diagram illustrating how a morphogen gradient acts on embryo cells.
(a) French flag model. Thresholding of the morphogen concentration gradient activate different genes in cells at different locations (indicated by different colors or shades). (b) Diffusing morphogens are absorbed by binding to cell surface receptors, resulting in the activation of downstream signaling pathways that switch various genes on or off. These may include the genes responsible for synthesis of the receptors themselves (nonlinear feedback).}
\label{mgrad}
\end{figure}

There are a number of important biological constraints on the effectiveness of any diffusion-based model of gradient formation. First, the concentration thresholding mechanism that determines the boundaries between differentiated cell populations has to be robust to both intrinsic and extrinsic noise fluctuations \cite{Eldar03,Barkai09,Howard12}. Extrinsic noise is usually associated with cell-to-cell variations in environmental factors such as rate of protein synthesis, whereas intrinsic noise refers to fluctuations within a cell due to biochemical reactions involving small numbers of molecules. Second, the rate of convergence to the steady-state concentration gradient should be consistent with the relevant biological time-scales.
One way to characterize the relaxation process is to treat the fractional deviation from the steady-state concentration as a cumulative distribution whose mean is identified with the so-called local accumulation time \cite{Berez10,Berez11,Gordon11}.

The classical diffusion-based mechanism for morphogen gradient formation can be interpreted as a model of diffusion in a one-dimensional (1D) domain $\Omega \subset \R$ with partial absorption at some constant rate $\kappa_0$ within some subdomain ${\mathcal U}\subseteq \Omega$. At the single particle level, this is a special case of a more general probabilistic framework for analyzing diffusion in partially absorbing media \cite{Bressloff22a,Bressloff22b}. The basic idea is to formulate the absorption process in terms of a generalized propagator $P(x,a,t)$, which is the joint probability density for particle position $X_t$ at time $t$ and the so-called occupation time $A_t$ in the absence of absorption. (This is a natural extension of the corresponding propagator for surface-based absorption, which involves the boundary local time \cite{Grebenkov20,Grebenkov22}.) The occupation time is a Brownian functional \cite{Majumdar05} that determines the amount of time that the particle spends within $\calU$. Partial absorption is then incorporated
 by introducing the stopping time 
$
{\mathcal T}=\inf\{t>0:\ A_t >\widehat{A}\}$,
 with $\widehat{A}$ a so-called stopping occupation time. Given the probability distribution $\Psi(a) = \P[\widehat{A}>a]$, the marginal probability density for particle position is defined according to
 $  p(x,t)=\int_0^{\infty} \Psi(a)P(x,a,t)da$. The classical example of partial absorption at a constant rate $\kappa_0$ corresponds to the exponential distribution
$\Psi(a) =\e^{-\kappa_0 a}$. This implies that one can obtain the generalized propagator $P(x,a,t)$ by Laplace transforming with respect to $a$, solving the resulting reaction-diffusion equation for a constant rate of absorption given by the corresponding Laplace variable $z$, and then calculating the inverse Laplace transform \cite{Bressloff22a,Bressloff22b}. Once the propagator has been determined, more general models of absorption with non-exponential distributions $\Psi(a)$ can be incorporated. These arise, for example, if the absorption rate depends on the occupation time. Indeed, a number of experiments suggest that various surface-based reactions are better modeled in terms of a reactivity that is a function of the surface interaction time \cite{Bartholomew01,Filoche08}. 
That is, the surface may need to be progressively
activated by repeated encounters with a diffusing
particle, or an initially highly reactive surface may become less active due to multiple interactions with the particle (passivation). Within the context of morphogen gradient formation, ``surface'' refers to the membrane of the differentiating cells.

We have recently shown that analogous macroscopic models of diffusion can be developed by reinterpreting the generalized propagator as a generalized concentration $C(x,a,t)$ with an associated marginal concentration $c(x,t)=\int_0^{\infty}\Psi(a) C(x,a,t)da$ \cite{Bressloff22c}.
In this paper we investigate the effects of non-exponential models of absorption on morphogen gradient formation by calculating the generalized concentration $C(x,a,t)$. The structure of the paper is as follows. In section 2 we briefly review the classical diffusion-based mechanism for gradient formation in a finite interval of length $L$ and define the local accumulation time. In section 3, we introduce the probabilistic model of single-particle diffusion in a partially absorbing medium, which is then used to develop the generalized diffusion-absorption model for gradient formation in section 4. We derive general expressions for the steady-state concentration and the accumulation time, which are written in terms of the inverse Laplace transform of the generalized concentration. In section 5, we evaluate these expressions in the limit $L\rightarrow \infty$ and show how modifying the absorption process can generate similar concentration profiles to the classical case, and yet significantly decrease the corresponding accumulation time. This is consistent with the idea that modifying the kinetics of absorption mainly affects the dynamical approach to steady-state, rather than the steady-state itself. Hence, without additional mechanisms, it does not enhance robustness to fluctuations in the rate of production.

\section{Classical diffusion-absorption model for gradient formation}

We begin by describing the simplest diffusion-based mechanism capable of generating a stationary concentration gradient in a 1D domain of length $L$. Suppose that a site of protein synthesis is localized at the end $x=0$. This generates a flux $J_0$ of morphogen particles that diffuse within the bulk domain and are subsequently absorbed at a constant rate $\kappa_0$ due to binding of morphogen to membrane-bound receptors. Let $c(x,t)$ denote the morphogen concentration at $x\in [0,L]$ and time $t$. The classical model for gradient formation takes $c(x,t)$ to evolve according to the reaction-diffusion equation
\begin{equation}
\label{grad0}
{\frac{\partial c}{\partial t}=D\frac{\partial^2c}{\partial x^2}-\kappa_0 c(x,t),\quad \left . -D\frac{\partial c}{\partial x}\right |_{x=0}=J_0,\quad \left . \frac{\partial c}{\partial x}\right |_{x=L}=0.}
\end{equation}
This has the steady-state solution
\begin{equation}
\label{ss0}
c^*(x)=\frac{J_0\cosh([L-x]/\lambda)}{D\lambda^{-1}\sinh(L/\lambda)},\quad \lambda=\sqrt{\frac{D}{\kappa_0}}.
\end{equation}
Note that when $L \ll \lambda $, the gradient is approximately linear, whereas when $\ L \gg \lambda $ it decays exponentially with length constant $\lambda$. In the latter case,
\begin{equation}
\label{grad}
c^*(x)=\frac{J_0\lambda}{D}\e^{-x/\lambda}.
\end{equation}

An important constraint on any model is that the time-scale of gradient formation is consistent with cellular time scales. A useful quantity for characterizing the time-dependent approach to steady-state of a diffusion process is the accumulation time \cite{Berez10,Berez11,Gordon11}. Consider the function
\begin{equation}
\label{accu}
Z(x,t)=1-\frac{c(x,t)}{c^*(x)},
\end{equation}
which represents the fractional deviation of the concentration from the steady-state. Assuming that there is no overshooting, $1-Z(x,t)$ is the fraction of the steady-state concentration that has accumulated at $x$ by time $t$. It follows that $-\partial_t Z(x,t)dt$ is the fraction accumulated in the interval $[t,t+dt]$. The accumulation time is then defined by analogy to mean first passage times 
\begin{equation}
\tau(x)=\int_0^{\infty} t\left (-\frac{\partial Z(x,t)}{\partial t}\right )dt=\int_0^{\infty} Z(x,t)dt.
\end{equation}
Note that a finite accumulation time implies that the steady-state $c^{*}(x)$ is a stable solution to (\ref{grad0}).
As a simple illustration of calculating $\tau(x)$, consider the the time-dependent solution of equation (\ref{grad0}) for $L\rightarrow \infty$:
\begin{eqnarray*}
\fl  c(x,t)&=&c^*(x)\left [1-\frac{1}{2}\mbox{erfc}\left (\frac{\sqrt{Dt}}{\lambda}-\frac{x}{2\sqrt{Dt}}\right )  -\frac{\e^{x/\lambda}}{2}\mbox{erfc}\left (\frac{\sqrt{Dt}}{\lambda}+\frac{x}{2\sqrt{Dt}}\right )\right ],
\end{eqnarray*}
where $\mbox{erfc}(z)$ is the complementary error function. It follows that
\begin{eqnarray*}
 Z(x,t)&=&\frac{1}{2}\mbox{erfc}\left (\frac{\sqrt{Dt}}{\lambda}-\frac{x}{2\sqrt{Dt}}\right )-\frac{\e^{x/\lambda}}{2}\mbox{erfc}\left (\frac{\sqrt{Dt}}{\lambda}+\frac{x}{2\sqrt{Dt}}\right ),
 \end{eqnarray*}
and \cite{Berez10}
\begin{equation}
\label{morpht}
\tau(x)=\frac{1}{2\kappa_0}(1+x/\lambda).
\end{equation}
For more complicated (linear) models it is often more convenient to work in Laplace space. Laplace transforming equation (\ref{accu}) with respect to $t$ and using the identity $c^*(x)=\lim_{s\rightarrow 0}s\widetilde{c}(x,s)$,
we have
\[s\widetilde{Z}(x,s)=1-\frac{s\widetilde{c}(x,s)}{c^*(x)}\]
and, hence
\begin{eqnarray}
\label{Tuc}
 \fl \tau(x)=\lim_{s\rightarrow 0} \widetilde{Z}(x,s) = \lim_{s\rightarrow 0}\frac{1}{s}\left [1-\frac{s\widetilde{c}(x,s)}{c^*(x)}\right ] =-\frac{1}{c^*(x)}
\left .\frac{d}{ds}s\widetilde{c}(x,s)\right |_{s=0}.
\end{eqnarray}

\section{Probabilistic model of diffusion in a partially absorbing medium}

The diffusion-absorption model (\ref{grad0}) can be interpreted as a model of diffusion in a 1D domain with partial absorption at a constant rate $\kappa_0$. In order to consider more general models of absorption, we turn to a probabilistic framework for analyzing single-particle diffusion in partially absorbing media \cite{Bressloff22a,Bressloff22b}. We formulate the absorption process in terms of a generalized propagator, which is the joint probability density for particle position $X_t$ and the occupation time $A_t$ in the absence of absorption. For the moment suppose that the particle is diffusing in $\Omega =[0,L]$ and we are interested in the amount of time it spends in the subinterval $\calM =[0,\ell]$, $0<\ell \leq L$. The associated occupation time is a Brownian functional \cite{Majumdar05} defined according to 
\begin{equation}
\label{occ}
A_t=\int_{0}^tI_{\calM}(X_{\tau})d\tau .
\end{equation}
Here $I_{\calM}(x)$ denotes the indicator function of the set $\calM$, that is, $I_{\calM}(x)=1$ if $x\in \calM$ and is zero otherwise. That is, $A_t$ specifies the amount of time the particle spends within $\calM$ over the time interval $[0,t]$. We also take $X_0=0$ and $A_0=0$.
Denoting the generalized propagator by $P(x,a,t )$ and $Q(x,a,t)$ for $x\in [\ell,L]$ and $x\in [0,\ell]$, respectively, we have the boundary value problem (BVP) \cite{Bressloff22a}
\numparts
\begin{eqnarray}
\label{Pocca}
\fl&\frac{\partial P(x,a,t)}{\partial t}=D\frac{\partial^2 P(x,a,t )}{\partial x^2}, \ x\in (\ell,L),\\
\fl &\frac{\partial Q(x,a,t )}{\partial t}=D\frac{\partial^2 Q(x,a,t )}{\partial x^2} -\left (\frac{\partial Q}{\partial a}(x,a,t ) +\delta(a)Q(x,0,t ) \right ),\ x\in (0,\ell),
\label{Poccb}\\
\fl &\left .\frac{\partial P(x,a,t )}{\partial x}\right |_{x=L}=0,\quad \left .\frac{\partial Q(x,a,t )}{\partial x}\right |_{x=0}=0.
\label{Poccc}
\end{eqnarray}
These are supplemented by matching conditions at the interface $x=\ell$,
	\begin{equation}
	\label{Poccd}
	\fl P(\ell,a,t )=Q(\ell,a,t ),\quad  \left .\frac{\partial P(x,a,t )}{\partial x}\right |_{x=\ell}=\left .\frac{\partial Q(x,a,t )}{\partial x}\right |_{x=\ell},
	\end{equation}
	\endnumparts
	and the initial conditions $P(x,a,0)=\delta(x)\delta(a)$, $Q(x,a,0)=0$.

A probabilistic model of partial absorption within $\calM$ can now be formulated as follows \cite{Bressloff22a}. Introduce the general stopping time condition
\begin{equation}
\label{TA}
{\mathcal T}=\inf\{t>0:\ A_t >\widehat{A}\},
\end{equation}
where $\widehat{A}$ is a random variable with probability distribution $\Psi(a)$. Heuristically speaking, ${\mathcal T}$ is a random variable that specifies the time of absorption in $\calM$, which is the event that $A_t$ first crosses a randomly generated threshold $\widehat{A}$. The marginal probability density for particle position $X_t $ is then 
\numparts
\begin{eqnarray}
\label{peep}
p(x, t)&=\int_0^{\infty}\Psi(a) P(x,a,t )da,\ x\in [\ell,L]\\
q(x,t)&=\int_0^{\infty}\Psi(a) Q(x,a,t )da,\ x\in [0,\ell].
\end{eqnarray}
\endnumparts
We now make the observation that if $\Psi(a)=\e^{-\kappa_0 a}$, then $p(x,t)=\PP(x,\kappa_0,t)$ and $q(x,t)=\Q(x,\kappa_0,t)$ with $\PP$ and $\Q$
satisfying the BVP obtained by Laplace transforming equations (\ref{Pocca})--(\ref{Poccd}) with respect to $a$ and identifying $\kappa_0$ as the Laplace variable:
\numparts
\begin{eqnarray}
\label{LTPocca}
&\frac{\partial \PP(x,\kappa_0,t)}{\partial t}=D\frac{\partial^2 \PP(x,\kappa_0,t)}{\partial x^2}, \ x\in (\ell,L)\\
&\frac{\partial \Q(x,\kappa_0,t)}{\partial t}=D\frac{\partial^2 \Q(x,\kappa_0,t)}{\partial x^2} -\kappa_0 \Q(x,\kappa_0,t) ,\ x\in (0,\ell),
\label{LTPoccb}\\
 &\left .\frac{\partial \PP(x,\kappa_0,t))}{\partial x}\right |_{x=L}=0,\quad \left .\frac{\partial \Q(x,\kappa_0,t)}{\partial x}\right |_{x=0}=0,
\label{LTPoccc}\\
 & \PP(\ell,\kappa_0,t)=\Q(\ell,\kappa_0,t),\quad  \left .\frac{\partial \PP(x,\kappa_0,t)}{\partial x}\right |_{x=\ell}=\left .\frac{\partial \Q(x,\kappa_0,t)}{\partial x}\right |_{x=\ell}.
\label{LTPoccd}
\end{eqnarray}
\endnumparts
Hence, single-particle diffusion in a partially absorbing subdomain $\calM$ with a constant rate of absorption $\kappa_0$ is equivalent to taking the distribution of the stopping occupation time to be an exponential, $\Psi(a)=\e^{-\kappa_0 a}$. (This is analogous to the equivalence of the classical Robin BVP and a partially absorbing surface with an exponential stopping local time distribution \cite{Grebenkov20}.) In other words, we can identify $\PP(x,\kappa_0,t)$ and $\Q(x,\kappa_0,t)$ as the marginal probability densities when the absorption rate is constant. 

The advantage of the propagator formalism is that a much wider class of absorption mechanisms can be modeled using different choices for $\Psi(a)$. For example, suppose the effective rate of absorption depends on the amount of time the particle spends within $\calM$, that is, $\kappa=\kappa(a)$. The corresponding stopping occupation time distribution takes the form
\begin{equation}
\Psi(a)=\exp \left (-\int_0^a\kappa(a')da'\right ).
\end{equation}
There are a variety of biophysical processes that could contribute to a non-constant rate of absorption, including changes in the conformational state of the particle, chemical reactions, and transient binding of the particle to subcellular substrates or buffers. Irrespective of the particular choice of distribution $\Psi(a)$, the marginal densities can be obtained by solving the Laplace transformed BVP (\ref{LTPocca})-(\ref{LTPoccd}), with $\kappa_0$ replaced by the general Laplace variable $z$, and then inverting with respect to $z$:
\begin{equation}
\label{pq}
\fl p(x,t)=\int_0^{\infty} \Psi(a){\mathcal L}_{a}^{-1}[\PP(x,z,t)] da,\ q(x,t)=\int_0^{\infty} \Psi(a){\mathcal L}_{a}^{-1}[\Q(x,z,t) ]da.
\end{equation}

\section{Generalized diffusion-absorption model for gradient formation}

The probabilistic model of single-particle diffusion presented in section 3 suggests a novel way to generalize the classical diffusion-absorption model of gradient formation. First, we identify the domains $\calM$ and $\Omega$ by setting $\ell=L$. Equations (\ref{Pocca})--(\ref{Poccd}) then reduce to the scalar BVP
\numparts
\begin{eqnarray}
\label{Qocca}
\fl &\frac{\partial Q(x,a,t )}{\partial t}=D\frac{\partial^2 Q(x,a,t )}{\partial x^2} -\left (\frac{\partial Q}{\partial a}(x,a,t ) +\delta(a)Q(x,0,t ) \right ),\ x\in (0,L),
\\
\fl & \left .\frac{\partial Q(x,a,t )}{\partial x}\right |_{x=0}=0=\left .\frac{\partial Q(x,a,t )}{\partial x}\right |_{x=L},\quad Q(x,a,0)=\delta(x)\delta(a).
\label{Qoccb}
\end{eqnarray}
\endnumparts
At first sight the introduction of the occupation time $A_t$ and associated propagator appears redundant, since $A_t=t$. However, the occupation time will allow us to incorporate absorption via the stopping time condition (\ref{TA}). Second, we consider a multiparticle version of the model in which the propagator $Q(x,a,t)$ is reinterpreted as the generalized concentration $C(x,a,t)$ of a large population of independently diffusing particles that are injected into the domain at different times so  $A_t\leq t$. One difference from the single particle BVP is that we can now include a source term at $x=0$:
\numparts
\begin{eqnarray}
\label{Cocca}
\fl &\frac{\partial C(x,a,t )}{\partial t}=D\frac{\partial^2 C(x,a,t )}{\partial x^2} -\left (\frac{\partial C}{\partial a}(x,a,t ) +\delta(a)C(x,0,t ) \right ),\ x\in (0,L)
\\
\fl & -D\left .\frac{\partial C(x,a,t )}{\partial x}\right |_{x=0}=J_0\delta(a),\quad \left .\frac{\partial C(x,a,t )}{\partial x}\right |_{x=L}=0.
\label{Coccb}
\end{eqnarray}
\endnumparts
In contrast to the single-particle case, we assume that the domain $[0,L]$ does not initially contain any particles so that $C(x,a,0)=0$. Finally, we introduce a model of absorption based on the stopping condition (\ref{TA}) so that at the multiparticle level, the concentration of activated proteins is
\begin{equation}
\label{cC}
c(x,t)=\int_0^{\infty} \Psi(a) C(x,a,t)da.
\end{equation}

Note that equations (\ref{Cocca}) and (\ref{Coccb}) can be rewritten as
\numparts
\begin{eqnarray}
\label{Cocca2}
\fl &\frac{\partial C(x,a,t )}{\partial t}+\frac{\partial C(x,a,t )}{\partial a}=D\frac{\partial^2 C(x,a,t )}{\partial x^2} ,\, x\in [0,L),
\\
\fl  & \left .\frac{\partial C(x,a,t )}{\partial x}\right |_{x=0}=0,\ a >0,\quad C(x,0,t ) =J_0\delta(x),\quad  \left .\frac{\partial C(x,a,t )}{\partial x}\right |_{x=L}=0.
\label{Coccb2}
\end{eqnarray}
\endnumparts
The sum of time derivatives on the left-hand side of equation (\ref{Cocca2}) reflects the fact that we have an age-structured model. Age-structured models are probably best known within the context of birth-death processes in ecology,
 where the birth and death rates depend on the age of the underlying populations 
\cite{McKendrick25,Foerster59,Chou16,Iannelli17}. These could be cells 
 undergoing differentiation or proliferation \cite{Zilman10,Iyer18}, or whole organisms undergoing 
reproduction \cite{Keyfitz05}. In our model the extra time variable is the occupation time rather than the age of a cell. The boundary condition at $a=0$ is obtained by noting that the flux condition $-D\partial_xC(0,a,t)=J_0 \delta(a)$ is equivalent to including the term $J_0\delta(a)\delta(x)$ on the right-hand side of equation (\ref{Cocca}).

\subsection{Steady-state concentration $c^*(x)$}

We will calculate the steady-state concentration $c^*(x)=\int_0^{\infty}\Psi(a)C^*(x,a)$ by solving the steady-state version of equations (\ref{Cocca2}) and (\ref{Coccb2}), which is given by
\numparts
\begin{eqnarray}
\label{ssCocca2}
\fl &\frac{\partial C^*(x,a )}{\partial a}=D\frac{\partial^2 C^*(x,a )}{\partial x^2} ,\, x\in [0,L),
\\
\fl  & \left .\frac{\partial C^*(x,a )}{\partial x}\right |_{x=0}=0,\ a >0,\quad C^*(x,0 ) =J_0\delta(x),\quad  \left .\frac{\partial C^*(x,a)}{\partial x}\right |_{x=L}=0.
\label{ssCoccb2}
\end{eqnarray}
\endnumparts
We proceed using separation of variables. Setting $C^*(x,a)=X(x)A(a)$ leads to the pair of ordinary differential equations
\begin{equation}
D\frac{d^2X}{dx^2}=-\lambda^2 X,\quad \frac{dA}{da}=\lambda^2 A.
\end{equation}
Imposing the boundary condition at $x=L$ yields a solution of the form $\cos(\lambda[L-x]/\sqrt{D})\e^{-\lambda^2 a}$. For $a>0$ we have $dX/dx=0$ at $x=0$, which 
implies that $\lambda=\lambda_n= n \pi \sqrt{D}/L$ for integers $n\geq 0$. This yields the general solution
\begin{equation}
C^*(x,a)=\sum_{n\geq 0} c_n \e^{-n^2\pi^2 Da/L^2}\cos(n\pi(L-x)/L).
\end{equation}
Finally, the coefficients $c_n$ are determined by the ``initial'' condition $C^*(x,0)=J_0\delta(x)$. Setting $a=0$ in the general solution gives
\begin{equation}
\fl C^*(x,0)=\sum_{n\geq 0} c_n  \cos(n\pi(L-x)/L)=\sum_{n\geq 0} c_n  (-1)^n \cos (n\pi x/L).
\end{equation}
Recall that one series representation of the Dirac delta function in $[0,L]$ is
\begin{equation}
\delta(x-x')=\frac{1}{L}\left (1+2\sum_{n\geq 0} \cos(n\pi x/L) \cos(n\pi x'/L)\right ).
\end{equation}
It follows that $c_0=J_0/L$ and $c_n=2(-1)^n J_0/L$ and thus
\begin{equation}
\label{Cstar}
\fl C^*(x,a)=\frac{J_0}{L}\left (1+2\sum_{n\geq 1}(-1)^n\e^{-n^2\pi^2 aD/L^2}\cos[n\pi  (L-x)/L] \right ).
\end{equation}
Finally, given a stopping occupation time distribution $\Psi(a)$, the steady-state concentration $c^*(x)$ is obtained from equation (\ref{cC}):
\begin{eqnarray}
\label{sscC}
\fl c^*(x)&=\int_0^{\infty} \Psi(a) C^*(x,a)da\nonumber \\
\fl &=\frac{J_0}{L}\left (\widetilde{\Psi}(0)+2\int_0^{\infty}\Psi(a)\sum_{n\geq 1}(-1)^n \e^{-n^2\pi^2 aD/L^2}\cos[n\pi  (L-x)/L]da \right ).
\end{eqnarray}
Equation (\ref{sscC}) implies that a necessary condition for the existence of a steady-state solution is $\widetilde{\Psi}(0)<\infty$. Moreover, using integration by parts, 
\[\fl \widetilde{\Psi}(0)=\int_0^{\infty} \Psi(a)da=[a\Psi(a)]_0^{\infty}-\int_0^{\infty}a\Psi'(a)da = \int_0^{\infty}a\psi(a)da=-\widetilde{\psi}'(0).\]
 Hence, $c^*(x)$ only exists if the stopping occupation time density $\psi(a)$ has a finite first moment, at least when $L <\infty$ (see section 5). An analogous result was previously obtained for diffusion in domains with partially absorbing boundaries \cite{Bressloff22c}.

An alternative method for computing $c^*(x)$ involves applying the double Laplace transform
\begin{equation}
 \label{dLT}
 \calC(x,z,s )\equiv \int_0^{\infty}\e^{-za}\int_0^{\infty}\e^{-st}C(x,a,t )dt\, da,
 \end{equation} 
to equations (\ref{Cocca}) and (\ref{Coccb}). This gives
\numparts
\begin{eqnarray}
&D\frac{\partial^2\calC(x,z,s)}{\partial x^2}-(s+z)\calC(x,z,s)=0,\ x\in (0,L)  
 \label{PlocLT}
 \\
& -D\left .\frac{\partial \calC(x,z,s)}{\partial x}\right |_{x=0}=\frac{J_0}{s} ,\quad \left .\frac{\partial \calC(x,z,s))}{\partial x}\right |_{x=L}=0.\end{eqnarray}
\endnumparts
The corresponding solution in Laplace space is straightforward to write down:
\begin{eqnarray}
\label{mQ1D}
 \calC(x,z,s)&= \frac{J_0\cosh(\alpha(s+z)[L-x])}{sD\alpha(s+z)\sinh(\alpha(s+z)L)} ,\quad \alpha(s)=\sqrt{\frac{s}{D}}.
\end{eqnarray}
Multiplying equation (\ref{mQ1D}) by $s$ and taking the limit $s\rightarrow 0$ yields
\begin{equation}
\label{squid}
\calC^*(x,z)=\frac{J_0\cosh(\alpha(z)[L-x])}{D\alpha(z)\sinh(\alpha(z)L)} .
\end{equation}
(This is equivalent to the solution (\ref{ss0}) for a constant rate of absorption $z=\kappa_0$ with $\lambda=1/\alpha(\kappa_0)$.)
It follows that the steady-state concentration $c^*(x)$ for a given stopping time distribution $\Psi(a)$ is
\begin{equation}
c^*(x)=\int_0^{\infty}\Psi(a){\mathcal L}_a^{-1}[\calC^*(x,z)]da.
\end{equation}
In order to find the inverse Laplace transforms we use the Bromwich integral
\begin{equation}
\label{brom1D}
C^*(x,a)=\frac{1}{2\pi i  }\frac{J_0}{D}\int_{\xi-i\infty}^{\xi+i\infty} \e^{za}\frac{\cosh[\alpha(z) (L-x)]}{\alpha(z)\sinh [\alpha(z)L]} dz. 
\end{equation}
Here $\xi$, $\xi>0$, is chosen so that the Bromwich contour is to the right of all singularities of $\calC^*(x,z)$. 
The Bromwich integral (\ref{brom1D}) can be evaluated by closing the contour in the complex $z$-plane. The resulting contour encloses a countably infinite number of poles, which correspond to the zeros of the function $\sinh [\alpha(z)L]$:
\begin{equation}
z=z_n \equiv -\frac{n^2\pi^2D}{L^2}. 
\end{equation}(The dependence of $\calC^*(x,z)$ on $\alpha(z)=\sqrt{z/D}$ suggests that $z=0$ is a branch point. However, $\cosh[\alpha(z) (L-x)]$ and $\alpha(z)\sinh [\alpha(z)L]$ are even functions of $\alpha(z)$, so $\calC^*(x,z)$ is actually single-valued.) 
Applying Cauchy's residue theorem to the Bromwich contour integral, and noting that the contribution from the semi-circle $C_R$ vanishes in the limit $R\rightarrow \infty$, we recover the solution (\ref{Cstar}).

\subsection{Accumulation time} One advantage of working in Laplace space is that we can also use the solution (\ref{mQ1D}) to calculate the accumulation time according to equation (\ref{Tuc})
with
\begin{eqnarray}
\fl \frac{d}{ds}s\widetilde{c}(x,s) &=\frac{d}{ds}\left (s\int_0^{\infty}\Psi(a){\mathcal L}_a^{-1}[\calC(x,z,s)]da\right )\nonumber \\
\fl &=\frac{J_0}{D}\frac{d}{ds}\left (\int_0^{\infty}\Psi(a){\mathcal L}_a^{-1}\left [ \frac{\cosh(\alpha(s+z)[L-x])}{\alpha(s+z)\sinh(\alpha(s+z)L)}\right  ]da\right ).
\end{eqnarray}
We will assume that the operations of differentiation, integration, and inversion of the Laplace transform all commute. We can then use the result
\begin{eqnarray}
\fl  \left .\frac{d}{ds}\frac{\cosh(\alpha [L-x])}{\alpha \sinh(\alpha L)}\right |_{s=0} &=\frac{1}{2 z\alpha(z)\sinh(\alpha(z)L)}\bigg \{ (L-x)\alpha(z)\sinh(\alpha(z)[L-x])\nonumber \\
\fl &\qquad -\left (1+L\alpha(z)\mbox{coth}(\alpha(z)L)\right )\cosh(\alpha(z)[L-x])\bigg \}\nonumber \\
\fl &\equiv -\Phi_L(x,z),
\label{PhiL}
\end{eqnarray}
and set
\begin{eqnarray}
\label{Tuc2}
  \tau(x)=\frac{1}{c^*(x)}\frac{J_0}{D}\int_0^{\infty}\Psi(a){\mathcal L}_a^{-1}[\Phi_L(x,z)]da
\end{eqnarray}
Note that $\tau(x)$ is independent of the flux $J_0$.

\section{Results}
In order to explore the effects of a non-constant reactivity, we ignore finite size effects by taking the limit $L\rightarrow \infty$. Equation (\ref{squid}) for the steady state $\calC^*(x,z)$ becomes
\begin{equation}
\calC^*(x,z)=\frac{J_0\e^{-\alpha(z)x}}{\sqrt{Dz}} .
\end{equation}
This can be inverted directly using standard Laplace transform tables to give
\begin{equation}
C^*(x,a)=J_0 \frac{\e^{-x^2/4Da}}{\sqrt{\pi a D}},
\end{equation}
and, hence,
\begin{equation}
\label{sJs}
c^*(x)=J_0\int_0^{\infty}\Psi(a) \frac{\e^{-x^2/4Da}}{\sqrt{\pi a D}}da.
\end{equation}
Similarly, from equation (\ref{PhiL}),
\begin{eqnarray}
\frac{J_0}{D}\lim_{L\rightarrow \infty}\Phi_L(x,z)&=\frac{J_0\e^{-\alpha(z)x}}{2z\sqrt{Dz}}[1+\alpha(z)x].
\end{eqnarray}
which can be inverted using
\begin{eqnarray}
{\mathcal L}_a^{-1} \left [\frac{\e^{-k\sqrt{z}}}{z}\right ]=\mbox{erfc}(k/2\sqrt{a}).
\end{eqnarray}
and
\begin{eqnarray}
{\mathcal L}_a^{-1} \left [\frac{\e^{-k\sqrt{z}}}{z^{3/2}}\right ]=2\sqrt{\frac{a}{\pi}}\e^{-k^2/4a}-k\, \mbox{erfc}(k/2\sqrt{a}).
\end{eqnarray}
Combining the various results leads to the following expression for the accumulation time:
\begin{eqnarray}
\tau(x)=\int_0^{\infty}
\Psi(a) \sqrt{\frac{a}{\pi}}\e^{-x^2/4Da}da \bigg \slash \int_0^{\infty}\Psi(a) \frac{\e^{-x^2/4Da}}{\sqrt{\pi a }}da.\label{acc1}
\end{eqnarray}

The functions $c^*(x)$ and $\tau(x)$ are well-defined provided that $c^*(0)$ and $\tau(0)$ are finite. This yields the integral conditions
\begin{equation}
\label{conn}
 \int_0^{\infty}\Psi(a) a^{\pm 1/2}da < \infty.
 \end{equation}
 These conditions can be related to certain constraints on
the moments of the stopping occupation time density $\psi(a)$. For example, if $\Psi(a)$ decays faster than $a^{-1/2}$ as $a \rightarrow \infty$, then
\begin{eqnarray}
c^*(0)&=\frac{J_0}{\sqrt{\pi D}}\int_0^{\infty}\Psi(a)a^{-1/2} da\nonumber \\
&=\frac{2J_0}{\sqrt{\pi D}}\left ([\Psi(a)a^{1/2} ]_0^{\infty} -\int_0^{\infty}\Psi'(a)a^{1/2} da\right )\nonumber\\
&=\frac{2J_0}{\sqrt{\pi D}}\int_0^{\infty}\psi(a)a^{1/2} da.
\end{eqnarray}
Similarly, if $\Psi(a)$ decays faster than $a^{-3/2}$ as $a \rightarrow \infty$, then
\begin{eqnarray}
\tau(0)&=\frac{1}{\sqrt{\pi} c^*(0)}\frac{J_0}{D} \int_0^{\infty}
\Psi(a) a^{1/2} da \nonumber \\
&=\frac{2}{3\sqrt{\pi} c^*(0)}\frac{J_0}{D}\left ([\Psi(a)a^{3/2} ]_0^{\infty} -\int_0^{\infty}\Psi'(a)a^{3/2} da\right )\nonumber\\
&=\frac{2}{3\sqrt{\pi} c^*(0)}\frac{J_0}{D} \int_0^{\infty}\psi(a)a^{3/2} da
\end{eqnarray}

\begin{figure}[b!]
\raggedleft
  \includegraphics[width=13cm]{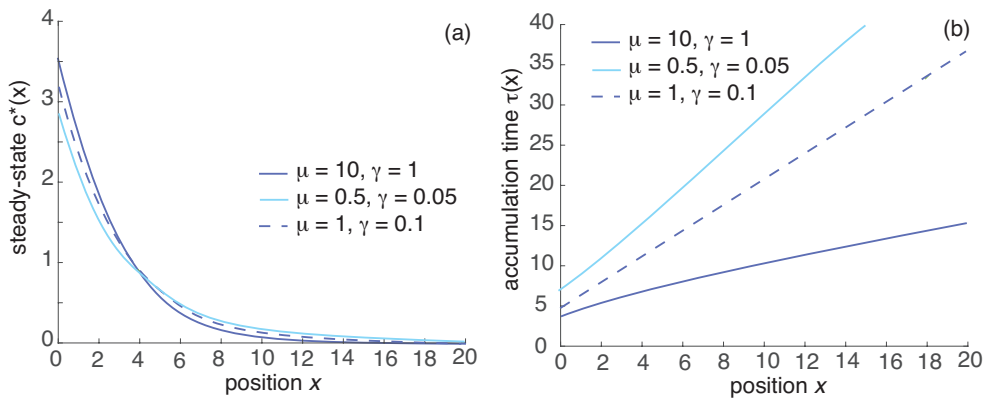}
  \caption{Gradient formation on the half-line for the gamma distribution. (a) Steady-state concentration profile $c^*(x)$ of equation (\ref{sJs}) and (b) accumulation time $\tau(x)$ of equation (\ref{acc1}) as functions of the position $x$ for different values of the parameters $(\mu,\gamma)$ with $\mu/\gamma$ fixed. Other parameters are $D=1,J_0=1$.}
  \label{fig2}
\end{figure}

An extensive list of common non-exponential distributions $\psi$ can be found in Ref. \cite{Grebenkov20}. Many of these examples have heavy tails and infinite first moments so that a steady-state concentration does not exist. One example of a non-exponential density that has finite moments and allows direct comparison with the standard exponential case, is the gamma distribution: 
\begin{equation}
\label{psigam}
\psi (a)=\frac{\gamma(\gamma a)^{\mu-1}\e^{-\gamma a}}{\Gamma(\mu)},\quad \Psi (a)=\frac{\Gamma(\mu,\gamma a)}{\Gamma(\mu)} ,\ \mu >0,
\end{equation}
where $\Gamma(\mu)$ is the gamma function and $\Gamma(\mu,z)$ is the upper incomplete gamma function:
\begin{equation}
\Gamma(\mu)=\int_0^{\infty}\e^{-t}t^{\mu-1}dt,\quad \Gamma(\mu,z)=\int_z^{\infty}\e^{-t}t^{\mu-1}dt,\ \mu >0.
\end{equation}
The corresponding reactivity is
\begin{equation}
 \kappa(a)=\gamma \frac{ (\gamma a)^{\mu-1}\e^{-\gamma a}}{\Gamma(\mu,\gamma a)}.
 \end{equation}
Here $\gamma$ determines the effective absorption rate so that the half-line $[0,\infty)$ is non-absorbing in the limit $\gamma\rightarrow 0$ and totally absorbing in the limit $\gamma \rightarrow \infty$. If $\mu=1$ then $\psi$ reduces to the exponential distribution with constant reactivity $\gamma$, that is, $\psi(a)|_{\mu =1}=\gamma \e^{-\gamma a}$. The parameter $\mu$ thus characterizes the deviation of $\psi(a)$ from the exponential case. If $\mu <1$ ($\mu>1$) then $\psi(a)$ decreases more rapidly (slowly) as a function of the occupation time $a$. In other words, the medium becomes more absorbing as $\mu$ decreases for fixed $\gamma$. These various features reflect the fact that $\E[a]=\mu/\gamma$. We can now determine the effects of a non-exponential distribution $\Psi(a)$ by fixing $\mu/\gamma=\mu'/\gamma'$ with $\mu =1$ and $\mu'\neq 1$. This is illustrated in Fig. \ref{fig2}, where we fix the spatial and temporal units by taking $D=J_0=1$. A major result of our analysis is that although the non-exponential case yields a similar steady-state concentration profile $c^*(x)$, there is a significant decrease (increase) in the accumulation time $\tau(x)$ as a function of $x$ when $\mu > 1$ ($\mu <1$). Moreover, $\tau(x)$ is a nonlinear function of $x$ when $\mu \neq 1$. This is consistent with the idea that modifying the kinetics of absorption mainly affects the dynamical approach to steady-state, rather than the steady-state itself. 

\begin{figure}[b!]
\raggedleft
  \includegraphics[width=13cm]{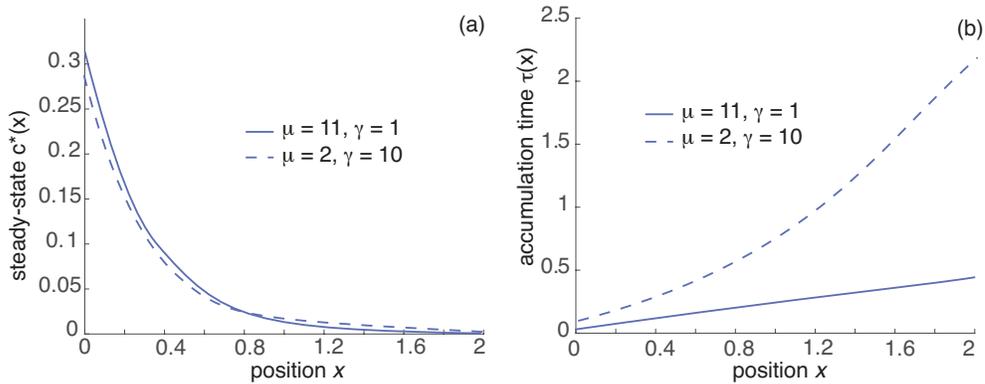}
  \caption{Gradient formation on the half-line for the Pareto-II distribution. (a) Steady-state concentration profile $c^*(x)$ of equation (\ref{sJs}) and (b) accumulation time $\tau(x)$ of equation (\ref{acc1}) as functions of the position $x$ for different values of the parameters $(\mu,\gamma)$ with $\gamma(\mu-1)$ fixed. Other parameters are $D=1,J_0=1$.}
  \label{fig3}
\end{figure}

Although we established our result for the gamma distribution, since it allows a direct comparison with standard models, analogous results hold for other distributions $\Psi(a)$, assuming that $c^*(x)$ and $\tau(x)$ exist and $\E(a)]$ is fixed. For example, consider the Pareto-II (Lomax) distribution
\begin{equation}
\psi(a)=\frac{\gamma \mu}{(1+\gamma a)^{1+\mu}},\quad  \Psi(a)=\frac{1}{(1+\gamma a)^{\mu}},\quad \mu >0.
\end{equation}
The corresponding reactivity is
\begin{equation}
\kappa(a)=\frac{\gamma \mu}{1+\gamma a}.
\end{equation}
In this case $\psi(a)$ only has finite moments when $\mu >1$; the blow up of the moments when $\mu<1$ reflects the fact that the Pareto-II distribution has a long tail. For example, the first moment is given by
\begin{equation}
\E[a]= \frac{1}{\gamma (\mu-1)}.
\end{equation}
In fact, the conditions (\ref{conn}) require that $\mu > 3/2$ for the Pareto-II distribution. In Fig. \ref{fig3} we compare two cases $(\mu,\gamma)$ and $(\mu',\gamma')$ with $\gamma (\mu-1)=\gamma'(\mu'-1)$. The concentration gradient is significantly shorter range than in the case of the gamma distribution. Again we find that similar gradients can have different accumulation time distributions. In conclusion, introducing a partial absorption scheme based on a two-parameter family of stopping occupation time distributions allows greater flexibility in jointly specifying the steady-state concentration gradient and the accumulation time.

\begin{figure}[t!]
  \raggedleft
    \includegraphics[width=8cm]{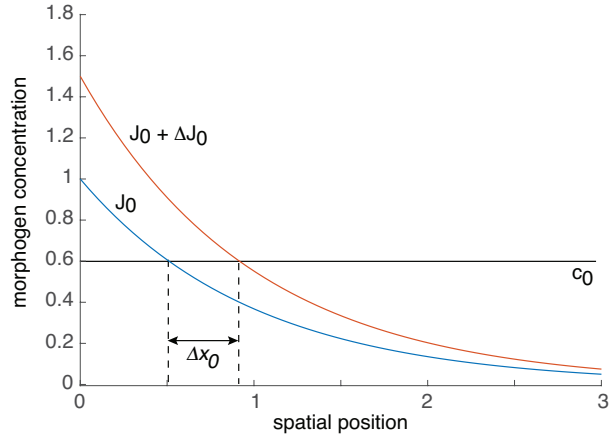}
  \caption{Effect of a shift in the flux, $J_0\rightarrow J_0+\Delta J_0$, on the position $x_0$ at which the gradient concentration drops below a critical level $c_0$.}
  \label{fig4}
\end{figure}

The above results also establishes that, without additional mechanisms, generalized absorption does not enhance robustness to fluctuations in the rate of production $J_0$.
A simple geometric argument is provided in Figure \ref{fig4}. First, rewrite the steady-state concentration as $c^*(x)=J_0\phi(x)$, with $\phi(x)$ independent of $J_0$. Let $x_0$ denote the position at which $c^*(x)=c_0$ for some fixed threshold $c_0$. That is, $x_0$ is the (unique) solution to the implicit equation
$\phi(x_0)=c_0/J_0$. It follows that $x_0=x_0(J_0)$ so under the fluctuation $J_0\rightarrow J_0+\Delta J_0$, there is a corresponding shift in the threshold crossing position, $x_0\rightarrow x_0+\Delta x_0$, with
\begin{equation}
x_0(J_0+\Delta J_0)=x_0(J_0)+\Delta x_0.
\end{equation}
Clearly, if the steady-state concentration has only a weak dependence on $\mu$ for fixed $\E[a]$, then the effective shift $\Delta x_0$ is also insensitive to $\mu$.

\section{Discussion} In this paper we showed how a more general probabilistic rule for partial absorption on a two-parameter family of stopping occupation time distributions $\Psi(a)$ allows greater flexibility in jointly specifying the steady-state concentration gradient and the accumulation time. In particular, one can maintain a similar range of morphogen signaling whilst significantly reducing the time to establish the gradient. Given that different choices for the distribution $\Psi(a)$ for fixed $\E[a]$ generate similar concentration profiles, one possible experimental method for distinguishing between the different models would be to determine the corresponding accumulation times. A further test of the models would be to check whether or not the accumulation time is independent of the rate of production $J_0$.

In future work it would be interesting to extend the theory to more complicated reaction-diffusion processes that include additional effects such as buffering \cite{Coppey08,Berez09}, and switching diffusivities \cite{Bressloff19}. Another possible generalization would be to consider absorption processes that depend nonlinearly on the protein concentration. It has been shown elsewhere that this provides a mechanism for increasing the robustness of gradient formation to environmental fluctuations  \cite{Eldar03}. However, the resulting nonlinearities complicate the analysis of gradient formation and the calculation of the accumulation time, for example \cite{Gordon11}. Within the context of the current paper, it would be necessary to develop a probabilistic formulation at the single particle level that yields a macroscopic diffusion equation that depends nonlinearly on the concentration; Laplace transform methods would no longer be applicable.

Finally, although concentration gradients are most familiar within the context of morphogenesis,
there is growing experimental evidence that concentration gradients also occur within individual cells. The existence of an intracellular gradient was first predicted theoretically by Brown and Kholodenko \cite{Brown99,Kholodenko09}, and has subsequently been found to play a role in a wide range of cellular processes, including cell division, polarity and mitotic spindle dynamics \cite{Kalb02,Caudron05,Meyers06,Garcia09,Garcia10}. An important difference between intracellular gradients and multicellular morphogen gradients is that absorption does not play a significant role in the formation of intracellular gradients. Instead, some modification in the protein, such as its phosphorylation state, changes as it moves away from the catalytic source of the modification. One thus has a gradient in the concentration of active protein. Mathematically speaking, the simplest model of intracellular gradient formation is identical in form to equation (\ref{grad0}) with $c(x,t)$ the concentration of active protein, $J_0$ the activation rate at one end, and $\kappa_0$ now a deactivation rate rather than an absorption rate. It would be interesting to identify deactivation mechanisms that also depend on the time a protein spends in contact with some reactive substrate. The machinery presented in this paper would then carry over to intracellular gradients.

\section*{References}

\begin{footnotesize}

\end{footnotesize}


\begin{thebibliography}{9}

\bibitem{Wolpert69} Wolpert L 1969 Positional information and the
spatial pattern of cellular differentiation. {\em J. Theor. Biol.}
{\bf 25} 1-47 

\bibitem{Wolpert06} Wolpert L. 2006 {\em Principles of Development}. Oxford, UK: Oxford Univ. Press.

\cite{Lander02,Ashe06,Wartlick09,Lander11,Shvartsman12,Teimouri16}. 

\bibitem{Lander02} Lander A D Nie W and Wan F Y 2002 Do morphogen gradients arise by diffusion? {\em Dev Cell }{\bf 2} 785-796  

  


\bibitem{Ashe06} Ashe H L and Briscoe J 2006 The interpretation of morphogen
gradients. {\em Development} {\bf 133} 385-394

\bibitem{Wartlick09} Wartlick O, Kicheva A and Gonzalez-Gaitan M 2009 Morphogen gradient formation. {\em Cold Spring Harb. Perspect. Biol.} {\bf 1} a001255  

\bibitem{Lander11} Lander A D 2011 Pattern, growth and control. {\em Cell} {\bf 144} 955-969

\bibitem{Shvartsman12} Shvartsman S Y and Baker R E 2012 Mathematical models of morphogen gradients and their effects on gene expression. {\em Rev. Dev Biol }{\bf 1} 715-730  

\bibitem{Teimouri16} Teimouri H and Kolomeisky A B 2016 Mechanisms of the formation of biological signaling profiles. {\em J. Phys. A: Math. Theor.} {\bf 49} 483001 

\bibitem{Eldar03} Eldar A, Rosin D, Shilo B-Z and Barkai N 2003 Self-enhanced ligand degradation underlies robustness of morphogen gradients. {\em Dev. Cell} {\bf 5} 635-646 

\bibitem{Barkai09} Barkai N and Shilo B Z 2009 Robust generation and decoding of morphogen gradients. {\em Cold Spring Harb. Perspect. Biol.}
{\bf 1} a001362  

\bibitem{Howard12} Howard M 2012 How to build a robust intracellular concentration gradient. {\em Trends Cell Biol.} {\bf 22} 311-317 (2012).


\bibitem{Berez10} Berezhkovskii A M, Sample C and Shvartsman S Y 2010 How
long does it take to establish a morphogen gradient?
{\em Biophys. J.} {\bf 99} L59-L61 


\bibitem{Berez11} Berezhkovskii A M, Sample C and Shvartsman S 2011 Formation
of morphogen gradients: local accumulation time.
{\em Phys Rev E} {\bf 83} 051906 

 \bibitem{Gordon11} Gordon P, Sample C, Berezhkovskii A M, Muratov C B and Shvartsman S 2011 Local kinetics of morphogen gradients.
{\em Proc Natl Acad Sci.}  {\bf 108} 6157-6162  


\bibitem{Bressloff22a} Bressloff P C 2022 Diffusion-mediated absorption by partially reactive targets: Brownian functionals and generalized propagators {\em J. Phys. A} {\bf 55} 205001.

\bibitem{Bressloff22b} Bressloff P C 2022 {The narrow capture problem: an encounter-based approach to partially reactive targets.} {\em Phys. Rev. E}. {\bf 105} 034141

\bibitem{Grebenkov20} Grebenkov D S 2020  {Paradigm shift in diffusion-mediated surface phenomena.} {\em Phys. Rev. Lett.} {\bf 125} 078102  

\bibitem{Grebenkov22} Grebenkov D S 2022  {An encounter-based approach for restricted diffusion with a gradient drift.}  {\em J. Phys. A}  {\bf 55} 045203

\bibitem{Majumdar05}  Majumdar S N 2005 Brownian functionals in physics and computer science. {\em Curr. Sci.} {\bf 89}, 2076  

\bibitem{Bartholomew01}Bartholomew CH. 2001 Mechanisms of catalyst deactivation,
{\em Appl. Catal. A: Gen.} {\bf 212}, 17-60 


\bibitem{Filoche08}  Filoche M, Grebenkov DS, Andrade Jr JS, 
Sapoval B 2008 Passivation of irregular surfaces accessed by
diffusion. {\em Proc. Natl. Acad. Sci.} {\bf 105}, 7636-7640

\bibitem{Bressloff22c} Bressloff P C 2022 Accumulation times for diffusion-mediated surface reactions. Submitted. arXiv:2205.08930 (2022).

\bibitem{McKendrick25} McKendrick A G 1925 Applications of mathematics to medical
problems. {\em Proc. Edinb. Math. Soc. } {\bf 44} 98  

\bibitem{Foerster59} Von Foerster H 1959 Some remarks on changing populations, in {\em The Kinetics of Cellular Proliferation.} edited by F.
Stohlman, Jr. Grune and Stratton, New York  



 \bibitem{Chou16} Chou T and Greenman C D 2016 A hierarchical kinetic theory of birth, death and fission in age-structured interacting populations {\em J. Stat. Phys.} {\bf 164} 49-76  
 
 \bibitem{Iannelli17} Iannelli M, and Milner F 2017 The basic approach to age-structured population dynamics: models, methods and numerics. {\em Lecture notes on mathematical modelling in the life sciences}. Springer
 
 \bibitem{Zilman10}  Zilman A, Ganusov V V and  Perelson A S 2010 Stochastic models of lymphocyte proliferation and death.
{\em PLoS One} {\bf 5} e12775 



\bibitem{Iyer18} Jafarpour F, Wright C S, Gudjonson H, Ridebling J, Dawson E, Lo K, Fiebig A, Crosson S, Dinner A R and Iyer-Biswas S 2018 Bridging the timescales of single-cell and population dynamics. {\em Phys. Rev. X} {\bf 8} 021007  

\bibitem{Keyfitz05} Keyfitz N and Caswell H 2005 {\em Appl. Math. Demogr.} 3rd edn. Springer, New York  


 
 \bibitem{Coppey08} Coppey M, Boettiger A N, Berezhkovskii A M and Shvartsman S Y 2008 Nuclear trapping shapes the terminal gradient
in the Drosophila embryo. {\em Curr. Biol.} {\bf 18} 915-919  

\bibitem{Berez09} Berezhkovskii A M, Coppey M and Shvartsman S Y 2009 Signaling gradients in cascades of two-state reaction-diffusion
systems. {\em Proc. Natl. Acad. Sci. USA} {\bf 106} 1087-1092 

 \bibitem{Bressloff19}  Bressloff P C, Lawley S C and Murphy P 2019 {Protein concentration gradients and switching diffusions.} {\em Phys. Rev. E} {\bf 99}  032409

\bibitem{Brown99} Brown G C and Kholodenko B N 1999 Spatial gradients of cellular phospho-proteins. {\em FEBS Lett.} {\bf 457} 452-454 

 \bibitem{Kholodenko09} Kholodenko B N 2009 Spatially distributed cell signalling. {\em FEBS Letters} {\bf 583} 4006-4012 


\bibitem{Kalb02} Kalab P, Weis K and Heald R 2002 Visualization of a {R}an-{GTP} gradient in
interphase and mitotic Xenopus egg extracts. {\em Science} {\bf 295} 2452-2456 

 \bibitem{Caudron05} Caudron M, Bunt , Bastiaens P and Karsenti E 2005 Spatial coordination of spindle assembly by chromosome-mediated signaling gradients. {\em Science} {\bf 309} 1373-1376 




 \bibitem{Meyers06} Meyers J, Craig J and Odde D J 2006 Potential for control of signaling pathways via cell size and shape. {\em Curr. Biol.} {\bf 16} 1685-1693 

\bibitem{Garcia09} Munoz-Garcia, J., Neufeld, Z., Kholodenko, B. N.: Positional information generated by spatially distributed signaling cascades. PLoS Comp. Biol. {\bf 3} e1000330 (2009).

 \bibitem{Garcia10} Munoz-Garcia J and Kholodenko B N 2010 Signaling and control from a systems perspective. {\em Biochem. Soc. Trans.} {\bf 38},1235-1241 (2010).
 
  
 






 
\end{thebibliography}
\end{document}